\documentclass [11pt]{article}

\usepackage{amsmath}
\usepackage{amsfonts}
\usepackage{amssymb}
\usepackage{ marvosym }
\newcommand{\Hom}{{\rm Hom }}
\newcommand{\conf}{Z_{{\rm conf}}}

\title{Electric-Magnetic Duality of Topological Gauge Theories from Compactification}

\author{     Ryan Thorngren
                       \thanks{Department of Mathematics, UC Berkeley.  ryan.thorngren@berkeley.edu}
       }

\begin{document}

\maketitle

\begin{abstract}  In this note, we discuss electric-magnetic duality between a pair of 4d topological field theories (TQFTs) by considering their compactifications to 2 dimensions. These TQFTs control the long-distance behavior of loop and surface operators in 4d gauge theories with gapped phases. These were recently used in work by S. Gukov and A. Kapustin in detecting phases not distinguishable by the Wilson-'t Hooft criterion and by A. Kapustin and the author to construct discrete theta-angles for lattice Yang-Mills theories. The strong-weak duality is manifested in an exchange of dynamical and background degrees of freedom in the compactified TQFTs. \end{abstract}

\section{Introduction}
The two theories that were discussed in \cite{GK} describe a Higgs and a confining phase where the unbroken gauge group and 't Hooft fluxes each take values in some finite group. Electric-magnetic duality in general should exchange these phases while dualizing the gauge group. We review these theories below.

What we will call the Higgs theory is defined by a high energy gauge group $G$ which is broken down to a finite abelian subgroup $\Gamma$. Writing $P$ for the principal $G$-bundle, there is a gauge field $A$, which is a connection on $P$, and the phase of a Higgs field $\phi$, which is a section of the associated bundle $P\times_G \mathfrak{h}$, where $\mathfrak{h}$ is the Lie algebra of the quotient $G/\Gamma = H$. There is also a Langrange-multiplier field $\lambda$, which is a 3-form valued in the dual space $P\times_G\mathfrak{h}^*$. We will also write $\mathfrak{g}$ for the Lie algebra of $G$ and $t:G\to H$ for the quotient map. In fact, $\mathfrak{g}$ and $\mathfrak{h}$ are isomorphic by the induced map $t_*$, but we keep the distinction explicit.

The action of this theory is a simple constraint,

\begin{equation}\label{higgsact}
\int (d\phi - t_*(A))\wedge \lambda.
\end{equation}

This action is manifestly topological. The 1-form wedging $\lambda$ is the covariant derivative $d_A\phi$ on the associated bundle where $\phi$ lives. The action thus enforces the constraint that $\phi$ is covariantly constant. This implies that $A$ is flat and has holonomies in $\text{\rm ker } t = \Gamma$. In fact, this theory has a lattice description as an untwisted Dijkgraaf-Witten theory for $\Gamma$.

This TQFT has unscreened Wilson loops with charges in $\hat\Gamma = \text{\rm Hom}(\Gamma, U(1))$ and unconfined magnetic surface operators with charges in $\Gamma$. For more on this TQFT's relationship to Higgs phases, we refer the reader to \cite{GK}.

The other theory is what we will call the confining theory. Its field content is slightly harder to describe mathematically. It has a high energy gauge group $G$ with a monopole condensate whose charges generate the finite normal subgroup $\Gamma$ of $G$. We will again write $H$ for the quotient $G/\Gamma$ and $t: G\to H$ for the quotient map. The field content includes a 2-form field $B$ valued in $\mathfrak{g}$. There is also a connection $A$ on a principal $H$-bundle and a Langrange-multiplier 2-form $\lambda$ valued in $\mathfrak{h}^*$.

It is worth discussing gauge transformations in the abelian case for 2-bundles. The gauge field $A$ has its ordinary gauge transformations, which don't affect $B$, but $B$ can be shifted by a connection $\alpha$ on an arbitrary principal $G$-bundle. This gauge transformation is

$$
A \mapsto A + t_*(\alpha)
$$
$$
B \mapsto B + d\alpha.
$$
Note that the topological class of the $A$ gauge bundle can change under this transformation. For more on gauge transformations, as well as the local description for the non-abelian case, see \cite{GK}.

The action for the confining theory is

\begin{equation}\label{confact}
\int (dA - t_*(B))\wedge \lambda.
\end{equation}
This action is also manifestly topological. The expression wedging $\lambda$ is something like a covariant derivative of $A$ with respect to $B$. The constraint $dA = t_*(B)$ implies a sort of flatness for $B$, where $B$'s holonomies around surfaces depend only on their homology class. For the abelian case this just means $dB=0$, but it is more complicated in general. It also forces these holonomies to be in $\text{\rm ker } t = \Gamma$. Indeed, in \cite{KT} it is shown that this theory has a formulation depending only on $\Gamma$.

This TQFT has unscreened 't Hooft loops with charges in $\Gamma$ as well as unconfined electric surface operators with charges in $\hat\Gamma$. For more on this TQFTs relation to confining phases, we refer the reader to \cite{GK} and \cite{KT}.

Note that a theory with this field content and action can also be defined if $t$ is not surjective but instead has finite cokernel. A detailed analysis of this case is forthcoming in \cite{KT2}.

The result, proven in \cite{KT} using lattice methods, which we explore in the continuous regime via compactification is that the Higgs theory defined by the data

$$
1\to \Gamma \to G \to H \to 1
$$
is dual to the confining theory associated to the dual data
$$
1\to \hat\Gamma \to \hat H \to \hat G \to 1,
$$
where $\hat G$ and $\hat H$ denote the Langlands-dual groups to $G$ and $H$, respectively. This is an instance of electric-magnetic duality. This duality does not hold in general when there are theta-angles; see \cite{KT}.

We analyze these theories by compactifying them. This takes some explanation, since they by nature don't depend on the metric. What we mean by compactification is that we consider quantities integrated over the compactified directions local in the lower dimensional effective theory. For more on this technique, see \cite{K}.

The paper is organized as follows. First we give a path integral argument for the duality in the abelian case. Then we use compactification to determine the partition function, vector space assigned to a 3-manifold, and category assigned to a Riemann surface. We conclude with a discussion of the duality from this perspective.

The author would like to thank Anton Kapustin, Alex Rasmussen, and Brian Swingle for useful discussions. This work is based on the author's SURF project in the Summer of 2012, and the author would like to thank the California Institute of Technology for the funding.

\section{Path Integral Proof of Duality in the Abelian Case}

In the abelian case, there is a path integral argument that the two theories are
dual. It suffices to consider the case $G=H=U(1)$ and $\Gamma = \mathbb{Z}_n$. The map $t$ is given by multiplication by $n$. We follow \cite{W}. 

One proceeds by introducing an auxiliary $H$ gauge
field $\hat A$ to the Higgs theory and adding a
Lagrange multiplier $B$ setting $\hat A$ to be trivial modulo
gauge transformations. The Lagrangian is thus modified to

$$
(d\phi - \hat A - n A) \lambda + dB \hat A.
$$
The 2-form field $B$ needs to be more general than a global 2-form for it to kill the holonomies of
$\hat A$ around non-trivial cycles. We can
integrate it out to obtain the Lagrangian for the Higgs theory. Instead of
integrating out $B$, however, we can gauge $\phi$ to zero and integrate
out $\lambda$ to obtain the Lagrangian

$$
  n B dA.
$$

Dualizing $A$ in a similar manner, we arrive at
the Lagrangian of the confining theory, proving that the abelian theories are dual. 

Note that in the non-abelian case, it is not clear how to make sense of terms
such as $B dA$, but this is a convincing argument for duality in general since it is shown in \cite{KT} that the theories only depend on $\Gamma$ anyway, so we can always use an abelian model. Also non-trivial theta-angles make integration of the fields not always possible.

\section{Codimensions 0,1, and 2 for the Higgs Theory}

The theories we've discussed so far are presented by Lagrangians, but since they are topological, there is another perspective one can take which is due to Atiyah \cite{A}. That is, these TQFTs should assign numbers to closed 4-manifolds, vector spaces to 3-manifolds, categories to 2-manifolds, and increasingly complicated algebraic objects to closed manifolds of higher codimension. These should all be compatible in the sense that this defines a functor from a certain 4-category of cobordisms to a $\mathbb{C}$-linear 4-category of vector spaces.

To summarize, the functor $Z_{\text{\rm Higgs}}$ sends
\begin{equation}\label{higgs0}
Z_{\text{\rm Higgs}}(\Sigma_4) = |\Gamma|^{b_1(\Sigma_4)-1}
\end{equation}\begin{equation}\label{higgs1}
Z_{\text{\rm Higgs}}(\Sigma_3) = \mathbb{C}[\text{\rm Hom}(H_1(\Sigma_3),\Gamma)]
\end{equation}\begin{equation}\label{higgs2}
Z_{\text{\rm Higgs}}(\Sigma_2) = \bigsqcup_{H^1(\Sigma_2,\Gamma)} \text{\rm Rep}_\Gamma.
\end{equation}

Throughout, the notation $\mathbb{C}[A]$ will mean the vector space with basis elements corresponding to elements of the set $A$. The rest of the data describe how the mapping class group acts in each case. For codimension 1, it acts through its action on the homology. In codimension 2 it is less clear, as we discuss at the end.

First, we calculate the partition function on a closed 4-manifold $\Sigma_4$. we use the Lorentz gauge $d\star A = 0$. The action then gets a new set of terms
$$
\chi_0 d \star A + \text{\rm ghost terms not involving $A$ or $\chi_0$,}
$$
where $\chi_0$ is a 0-form Lagrange multiplier field valued in $\mathfrak{g}^*$. Since the induced map $t_*:\mathfrak{g}\to\mathfrak{h}$ is an isomorphism on, one can change variables, instead integrating over the $\mathfrak{h}$ valued forms $t_*A$. One should also change variables and integrate over $t_*^{\dagger -1} \chi_0$, where $t_*^\dagger:\mathfrak{h}^*\to\mathfrak{g}^*$ is the adjoint.

We're going to need to take determinants of these maps, so we should use compatible Killing forms on $G$ and $H$ and use these to identify $\mathfrak{g}$ and $\mathfrak{g}^*$, $\mathfrak{h}$ and $\mathfrak{h}^*$. Then it is easy to check that the determinants are simply the number of sheets of the cover: $\text{\rm det }t_* = |\Gamma|$, $\text{\rm det }t_*^{\dagger -1} = \text{\rm det }t_*^{-1} = |\Gamma|^{-1}$.

The resulting action is that for the theory with $G=H$ and $t={{\rm id}}$. The only contribution is thus from how the path integral measure transforms under this change of variables. Using the determinants calculated above, we have $\mathcal{D}A = |\Gamma|^{b_1 - B_1} \mathcal{D} t_* A $ and $\mathcal{D}\chi_0 = |\Gamma|^{B_0 - b_0} \mathcal{D} t_*^\dagger \chi_0$, where $b_k$ is the $k$th Betti number, and $B_k$ is the ``number" of k-forms. The Betti numbers appear because they are the dimension of the space of harmonic forms, which are the zero modes for this action, and we must remove them when considering the determinants in the path integral measure. The $B_k$ give cut-off dependent terms, so we may discard them. Noting $b_0 = 1$ for connected $\Sigma_4$, we thus obtain the result \eqref{higgs0}. The cut-off dependent terms are discussed in greater detail in a similar context in \cite{W}.

To calculate the Hilbert space assigned to a closed 3-manifold $\Sigma_3$, we compactify the theory on $\Sigma_3$ to obtain a 0+1 dimensional theory, whose Hilbert space is that which we seek. This theory has only finitely many configurations, so this space will be a sum of 1-dimensional vector spaces for each one, which are in bijection with $\Hom(H_1(\Sigma_3),\Gamma)$. We thus obtain the result \eqref{higgs1}.

The other important datum in codimension 1 is the action of the mapping class group of $\Sigma_3$. This has an evident action on the space of configurations via its action on $H_1(\Sigma_3)$, and there is no source of phase factors in the data of this theory to complicate this action, so this must be it.

We can check this answer has at least the right dimension by computing the partition function again. That is, $Z_{{\rm Higgs}}(\Sigma_3 \times S^1) = {{\rm dim }}Z_{{\rm Higgs}}(\Sigma_3) = |\Gamma|^{b_1(\Sigma_3)}$, and $b_1(\Sigma_3\times S^1) = b_1(\Sigma_3) + 1$, in agreement with the calculation above.

We can check the action of the mapping class group in a special case as well. Consider $\Sigma_3$ a 3-torus, and glue $\Sigma_4$ from $\Sigma_3$ by exchanging the longitude and latitude of a 2-torus factor of $\Sigma_3$. The result is $\Sigma_4 = S^1 \times X^3$, where $X^3$ is the 3-sphere with the two components of the Hopf link identified. This space has $H_1(X^3) = \mathbb{Z}$ corresponding to a segment from one component to the other, so $b_1(\Sigma_4)-1 = 1$. Meanwhile, the trace of the corresponding operator on the Hilbert space is $|\Gamma|$, so we see that the two calculations agree.

To calculate the category assigned to the Riemann surface $\Sigma_2$, consider another Riemann surface $X$ and $\Sigma_4 = \Sigma_2 \times X$. This 4-manifold is necessarily torsion-free. This allows us to write $A = A_{1,0} + A_{0,1}$, with $A_{1,0}$ locally a 1-form along $\Sigma_2$ and a 0-form along $X$, $A_{0,1}$ a 0-form along $\Sigma_2$ and locally a 1-form along $X$ and likewise for the Lagrange multiplier, $\lambda$. We use this notation throughout. The Lagrangian is then
$$
(d_2\phi - t_*A_{0,1})\lambda_{1,2} + (d_X\phi - t_*(A_{0,1}))\lambda_{2,1},
$$
where $d_2$ is the covariant derivative along $\Sigma_2$ and $d_X$ is that along $X$. Integrating over $\Sigma_2$, we see that the resulting theory is a direct sum of Higgs theories on $X$ labeled by background $\Gamma$-connections on $\Sigma_2$, ie. by $\Hom(H_1(\Sigma_2),\Gamma)=H^1(\Sigma_2,\Gamma)$.

Boundary conditions for the 2d Higgs theory are given by Wilson lines, so for each of these theories the category is ${{\rm Rep}}_\Gamma$, the category of complex representations of $\Gamma$. The whole category is the disjoint union of these for each direct summand.

We can check this answer with the previous analysis. It is easily seen that the compactification gives the correct Hilbert space and partition function for $\Sigma_2\times S^1$ and $\Sigma_2 \times S^1 \times S^1$, respectively.

We can also see this directly from the category. The Hilbert space assigned to $\Sigma_2\times S^1$ should be the ``dimension" of this category, the vector space of natural transformations from the identity functor to itself. Each ${{\rm Rep}}_\Gamma$ component contributes $\mathbb{C}[\Gamma]$, and there are $\Hom(H_1(\Sigma_2),\Gamma)$ of these, so the Hilbert space is $\mathbb{C}[\Hom(H_1(\Sigma_2)\oplus \Gamma , \Gamma)]$ = $\mathbb{C}[\Hom(H_1(\Sigma_2\times S^1),\Gamma)]$, in agreement with our previous calculations.

\section{Codim 0,1, and 2 for the Confining Theory}

We perform the same analysis for the confining theory. To summarize, we find
\begin{equation}\label{conf0}
\conf(\Sigma_4) = |\Gamma|^{b_2(\Sigma_4)-b_1(\Sigma_4)+1}
\end{equation}\begin{equation}\label{conf1}
\conf(\Sigma_3) = \mathbb{C}[H^2(\Sigma_3,\Gamma)]
\end{equation}\begin{equation}\label{conf2}
\conf(\Sigma_2) = \bigsqcup_\Gamma\bigotimes_{b_1(\Sigma_2)} \text{\rm Rep}_\Gamma,
\end{equation}
with the action of the mapping class group in codimension 1 acting through its action on cohomology. As with the Higgs theory, the action is unclear in codimension 2. We discuss this at the end.

To calculate the partition function on a closed 4-manifold $\Sigma_4$, we use the Lorentz gauge $d\star B = 0$. This requires a two-stage BRST action, where the $\mathfrak{g}^*$-valued Lagrange multiplier $\pi_1$ needs to be gauge fixed as well. This is given by
$$
\pi_1 d \star B + E_0 d \star \pi_1 + {\text{\rm ghost  terms not involving $B$, $\pi_1$, or $E_0$}}.
$$
We perform a change of variables as before, instead integrating over $t_*B$, $t_*^{\dagger-1}\pi_1$, and $t_*E_0$. 

The resulting action is that for a confining theory with $G=H$ and $t={{\rm id}}$, which describes a trivial theory, so again the only contribution is from the change of measure in the path integral.. Once again, zero modes are harmonic forms, so the path integral measure transforms by $|\Gamma|^{b_2(\Sigma_4)-b_1(\Sigma_4) +1}$, as well as some cut-off dependent terms we discard. Since the remaining integral is the partition function of a trivial theory, we can normalize it to be just $1$. This gives the answer quoted above in \eqref{conf0}.

The Hilbert space assigned to a closed 3-manifold $\Sigma_3$ will again be a sum of 1-dimensional vector spaces for each of the finitely many vacuum configurations on $\Sigma_3 \times \mathbb{R}$, which is homotopy equivalent to $\Sigma_3$.

 There is some subtlety to these configurations when $\Sigma_3$ has torsion. To see this, first consider the abelian case $H=G=U(1)$, $t$ is multiplication by $n$, and so $\Gamma = \mathbb{Z}_n$. The 1-form gauge transformations are
$$
B \mapsto B + d\xi
$$
$$
A \mapsto A + n\xi,
$$
where $\xi$ is an arbitrary $U(1)$ connection.

The equations of motion imply that $B$ defines a homomorphism $H_2(\Sigma_3) \to \mathbb{Z}_n$. This homomorphism determines $B$ up to gauge transformations.

It remains to see what data $A$ contributes to the configuration. Once we fix a representative for $B$, the remaining gauge transformations are those determined by flat connections $\xi$. These are determined by their holonomy morphism $H_1(\Sigma_3) \to U(1)$. We see that we can use this to cancel all but the $n$-torsion holonomy of $A$. Thus, the vacuum data sits in a short exact sequence
$$
\{\text{\rm $A$ data of $n$ torsion holonomy}\} \to \{ \text{\rm vacuum data} \} \to \{ \text{\rm $B$ holonomy data} \}.
$$
In fact, this is the universal coefficient sequence
$$
{{\rm Ext}}(H_1(\Sigma_3),\mathbb{Z}_n) \to H^2(\Sigma_3,\mathbb{Z}_n) \to \Hom(H_2(\Sigma_3),\mathbb{Z}_n),
$$
so the vacuum configurations are cohomology classes in $H^2(\Sigma_3,\mathbb{Z}_n)$.

The non-abelian case is harder to consider this way because the gauge transformations are much more involved. However, as explained in \cite{KT}, the answer only depends on $\Gamma$. Thus we obtain the Hilbert space $\conf(\Sigma_3) = \mathbb{C}[H^2(\Sigma_3,\Gamma)]$. It is easy to check that the dimension of this is $\conf(\Sigma_3\times S^1)$.

The mapping class group of $\Sigma_3$ permutes these configurations via its action on the cohomology group, and there is no source of phases that could complicate this action, so it acts by permutation also on the Hilbert space.

To go to codimension 2, we will use the abelian notation for simplicity, but, as before, the same results hold in the non-abelian case.

On $\Sigma_4 = \Sigma_2 \times X$, as in the Higgs case, we can split the fields so the Lagrangian is
$$
(d_2A_{1,0} - nB_{2,0})\lambda_{0,2} + (d_2 A_{0,1} + d_X A_{1,0} - n B_{1,1})\lambda_{1,1} + (d_X A_{0,1} - nB_{0,2})\lambda_{2,0}.
$$
A connection $\alpha = \alpha_{1,0} + \alpha_{0,1}$ defines a gauge transformation
$$
A_{1,0} \mapsto A_{1,0} + n\alpha_{1,0}
$$
$$
B_{2,0} \mapsto B_{2,0} + d_2\alpha_{1,0}
$$
$$
B_{1,1} \mapsto B_{1,1} + d_2\alpha_{0,1} + d_X\alpha_{1,0},
$$
and likewise for $B_{0,2}$ and $A_{0,1}$. 

For any 1-cycle $\gamma$ in $\Sigma_2$, we obtain a gauge field $C_\gamma$ on $X$ by integrating $B_{1,1}$ over $\gamma$. Under a gauge transformation, it becomes
$$
\delta C_\gamma = \int_\gamma d_X\alpha_{1,0} + d_2\alpha_{0,1} = d_X \int_\gamma \alpha_{1,0}.
$$
The integral on the right is a scalar field on $X$, so this is indeed an ordinary gauge transformation. Though these fields are nonlocal in the 4d theory, we consider them local in the effective 2d theory on $X$ obtained by compactifying $\Sigma_2$. We will use them to define boundary conditions for this theory.

Consider a 2-chain $\Lambda$ in $\Sigma_2$ with $\partial \Lambda = \gamma - \gamma'$. Then 
$$
C_\gamma - C_{\gamma'} = \int_\Lambda d_2 B_{1,1}.
$$
The equations of motion imply $dB=0$ and hence $d_2B_{1,1} = -d_X B_{2,0}$, so on-shell,
$$
C_\gamma - C_{\gamma'} = -d_X \int_\Lambda B_{2,0},
$$
which is a gauge transformation. Thus, if we want to consider boundary conditions, we can consider $\gamma$ to be a homology class an take $C_\gamma$ defined with respect to some representative.

The middle term in the action above sets $C_\gamma$ to be flat and have holonomy in $\mathbb{Z}_n$. Identifying $C_\gamma$ and $C_{\gamma'}$ for homologous 1-cycles $\gamma$ and $\gamma'$, we thus obtain $b_1(\Sigma_2)$ copies of the 2d Higgs theory. The rest of the action is composed of a 2d confining theory (which is in fact trivial) and a background $\mathbb{Z}_n$ flux through $\Sigma_2$.

The partition function as calculated by this decomposition is
$$
n^{b_1(\Sigma_2)b_1(X)-b_1(\Sigma_2)+2-b_1(X)+1} = n^{b_2(\Sigma_2\times X) - b_1(\Sigma_2 \times X) +1},
$$
in agreement with equation \eqref{conf0}.

For $X = S^1 \times \mathbb{R}$, there are $n$ sectors for the background flux through $\Sigma_2$, in which the $b_1(\Sigma_2)$ gauge fields each contribute $n$ vacua, while the $B$ field on $X$ doesn't contribute any nontrivial configurations since $b_2(X) = 0$. Thus, the Hilbert space is
$$
\conf(\Sigma_2\times S^1) = \mathbb{C}[\mathbb{Z}_n^{b_1(\Sigma_2)+1}] = \mathbb{C}[\mathbb{Z}_n^{b_1(\Sigma_2\times S^1)}],
$$
also in agreement with equation \eqref{conf1}.

Only the gauge fields can be used to define boundary conditions for the effective 2d theory on $X$. These are given by Wilson loops. The category of boundary conditions is thus
$$
\conf(\Sigma_2) = \bigsqcup_{\mathbb{Z}_n} \bigotimes_{b_1(\Sigma_2)} {{\rm Rep}}_{\mathbb{Z}_n},
$$
where the tensor product denotes the Deligne product. For general $\Gamma$ we have the result in equation \eqref{conf2}. Just as in the Higgs case, one can verify that this category has the right dimension.

\section{Duality and Concluding Remarks}

From the results argued above, we see some superficial disagreement between the two theories. For one the partition functions $\eqref{higgs0}$ and $\eqref{conf0}$ differ by a factor $|\Gamma|^{\chi(\Sigma_4)}$. However, such a factor can be obtained by adding a curvature dependent term to the action, and topological theories should be considered only modulo such terms.

The Hilbert spaces $\eqref{higgs1}$ and $\eqref{conf1}$ for $\Gamma$ and its Pontryagin dual are isomorphic by Poincar\'e duality. In detail, $H^2(\Sigma_3,\Gamma)$ is naturally isomorphic to $\Hom(H_1(\Sigma_2),\hat\Gamma)$, and the action of the mapping class group on the Hilbert spaces is via its action on these groups.

The question of how the categories \eqref{higgs2} and \eqref{conf2} are isomorphic is more subtle. They have the correct dimension, but it is unclear how the mapping class group could act on them. While a cohomology group appears in $\eqref{higgs2}$, for the confining theory there is no obvious topological index that the mapping class group could permute. We leave the solution to this puzzle to later work and perhaps to a different approach.

Another curious aspect of the duality is that in compactifying we see certain degrees of freedom become background fluxes, while others remain dynamical data. For instance, in the 2d compactification of the confining theory, we saw a background $B$-flux and dynamical gauge fields, while the 2d compactification of the Higgs theory had background electric fluxes and a single dynamical gauge field. The dynamical and background data appear to be interchanged by the duality map. This is a manifestation of the strong-weak coupling exchange.

\end{document}